\documentclass[aps,pra,twocolumn,showpacs]{revtex4-1}

\usepackage{braket,mathtools,color,pgfplots,graphicx}
\usepackage[colorlinks=true, citecolor=blue, filecolor=green, urlcolor=blue]{hyperref}
\usepackage[caption=false]{subfig}

\begin{document}
	
\title{Analytic Few Photon Scattering in Waveguide QED}
\author{David L. Hurst}
\email{dhurst1@sheffield.ac.uk}
\affiliation{Department of Physics and Astronomy, University of Sheffield, Hounsfield Road, Sheffield, S3 7RH, United Kingdom}
\author{Pieter Kok}
\email{p.kok@sheffield.ac.uk}
\affiliation{Department of Physics and Astronomy, University of Sheffield, Hounsfield Road, Sheffield, S3 7RH, United Kingdom}
\date{\today}
	
\begin{abstract}
\noindent We develop an approach to light-matter coupling in waveguide QED based upon scattering amplitudes evaluated via Dyson series. For optical states containing more than single photons, terms in this series become increasingly complex and we provide a diagrammatic recipe for their evaluation, which is capable of yielding analytic results. Our method fully specifies a combined emitter-optical state that permits investigation of light-matter entanglement generation protocols. We use our expressions to study two-photon scattering from a $\Lambda$-system and find that the pole structure of the transition amplitude is dramatically altered as the two ground states are tuned from degeneracy.
\end{abstract}
	
\maketitle

\section{Introduction}
\noindent
Proposals for devices such as a measurement-based quantum computer \cite{PhysRevLett.86.5188} or a Quantum Internet \cite{cite-key} require large entangled states with many stationary qubit nodes. Optical photons, with their long coherence times and large velocities, form the ideal carriers of quantum information between these nodes \cite{RevModPhys.87.1379} and this means that understanding the light-matter interaction is necessary for the purposes of practical device design \cite{PhysRevLett.117.240501}. A possible route towards engineering this light-matter interaction involves coupling quantum emitters to the modes of a nanophotonic waveguide. Recently, there has been a great deal of theoretical interest \cite{PhysRevA.94.043839,2017arXiv170501967F} and experimental progress in this field \cite{Sipahigilaah6875,cite-keyrev,cite-keyexp,ncomms}.

Photon scattering from a waveguide embedded emitter is a well-studied problem, with recent developments including the single and multi-photon scattering matrix \cite{PhysRevA.82.063821,PhysRevA.91.043845} and generalisations of the input-output formalism and master equation \cite{1367-2630-17-11-113001} to waveguide systems. There has also been a substantial body of work focussed on applying techniques from relativistic quantum field theory to the problem, notably the LSZ reduction formula \cite{PhysRevB.79.205111}, cluster decomposition principle \cite{PhysRevLett.111.223602} and diagrammatic evaluation of Green's functions \cite{PhysRevA.95.053845}. Interest in this simple system remains high today \cite{PhysRevA.93.063807,2017arXiv170104709S}, with many authors noting also the possibility of engineering strong on-chip photon-photon interaction \cite{1367-2630-17-2-023030}. 

Schemes for engineering entanglement between matter qubits \cite{PhysRevA.71.060310,PhysRevB.88.245306} require the stationary qubit state conditional on that of the optical field. This is not a universal feature of previously reported techniques, although is rendered possible by some very recent works \cite{PhysRevA.92.053834,1367-2630-18-9-093035}. We develop a formalism that fully specifies the combined emitter-optical state following photon scattering from a waveguide-embedded emitter. It is interesting to note that despite the apparent simplicity of the system, many of the previous approaches involve extremely advanced mathematical techniques and tend not to encourage an intuitive understanding of the global dynamics. This is something that the method we develop here avoids, with terms in each expression corresponding very naturally to physical processes. We can use this to visualise the processes not allowed by our initial choice of system Hamiltonian. 

In this paper, we consider multi-photon scattering from a waveguide embedded emitter and derive a method to determine analytic expressions for the transition amplitude between arbitrary combined emitter-optical input and output states. In Sec. \ref{sec:setup} we describe the system under analysis and outline a general procedure for specifying the global dynamics. In Sec. \ref{sec:tls} we perform this procedure explicitly for the case of a single two-level-system (TLS) with one and two-photon optical inputs. In Sec. \ref{sec:lambda} we demonstrate how to extend the developed diagrammatic approach for the scenario where the TLS is replaced with a $\Lambda$-system. This allows us, in Sec.~\ref{sec:entangle}, to study the pole structure of the transition amplitude for both cases. Sec.~\ref{sec:conc} is reserved for a summary, conclusions and some suggestions for future work.

\section{Waveguide QED System}
\label{sec:setup}
\begin{figure}[b!]
	\begin{center}
		\includegraphics[scale=1]{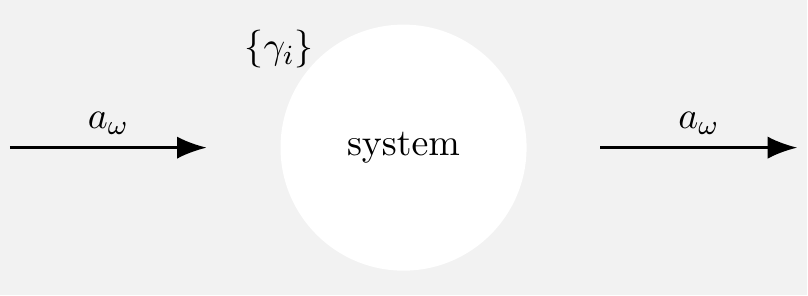}
		\caption{Some unspecified local system interacting with the continuum of optical bosonic modes $a_\omega$. The system may be composed of several sub-systems and thus the coupling is characterised in general by the set of rates $\{\gamma_i\}$. }
		\label{fig:fig1}
	\end{center}
\end{figure}
\noindent
The system analysed in this work consists of some general local system chirally coupled to the right-propagating modes $a_\omega$ of an optical waveguide \cite{cite-keyrev,cite-keyexp,ncomms}. The system is in general complex and composed of multiple sub-systems; therefore the coupling is characterised by the set of numbers $\{\gamma_i\}$. At some time $t_i\rightarrow-\infty$ the system state is given by $\ket{\phi_\text{in};\psi_\text{in}}$, where $\psi_\text{in}$ represents the optical wavefunction and $\ket{\phi_\text{in}}$ is the state in which the emitter is prepared. A scattering event then occurs, and the global system dynamics are in general complicated to describe until a time $t_f\rightarrow+\infty$, when the emitter has relaxed to some ground or meta-stable level and the optical state $\ket{\psi_\text{out}}$ is coupled out of the waveguide. Working in the interaction picture, the input and output states are eigenstates of the free Hamiltonian ($H_0$) that describes the dynamics of an uncoupled waveguide-emitter system \cite{schwartz2014quantum}. This allows us to construct input and output optical states from the usual creation and annihilation operators for photons. The transition amplitude $\mathcal{A}\equiv\bra{\phi_\text{out};\psi_\text{out}}\mathcal{U}\ket{\phi_\text{in};\psi_\text{in}}$ gives the overlap between an output state $\ket{\phi_\text{out};\psi_\text{out}}$ and an input state evolved from $t\rightarrow-\infty$ to $t\rightarrow+\infty$ by the operator $\mathcal{U}$. When $\mathcal{U}$ is the time evolution operator evaluated in the interaction picture, which in the long time limit is equivalent to the scattering matrix of the system, the transition amplitude $\mathcal{A}$ completely specifies the global system dynamics. 
\begin{figure}[t!]
	\begin{center}
		\includegraphics[scale=1]{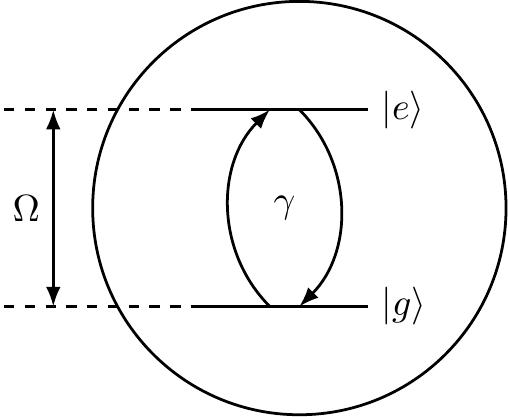}
		\caption{A two-level system. The excited state $\ket{e}$ is separated from the ground level $\ket{g}$ by the energy gap $\Omega$ and the two states are coupled with strength $\gamma$.   }
		\label{fig:TLS}
	\end{center}
\end{figure}
The expansion of $\mathcal{U}$ is known as the Dyson series \cite{PhysRevA.90.063840} and takes as an argument the global system interaction Hamiltonian $H_\text{I}(t)$. This operator is defined by ${H_\text{I}(t)\equiv e^{iH_0t}H_\text{int}e^{-iH_0t}}$, where $H_0$ is the free Hamiltonian of the system, $H_\text{int}$ the interaction Hamiltonian in the Shr\"{o}dinger picture and $\hbar=1$. We expand the transition amplitude $\mathcal{A}\equiv\mathcal{A}^{(0)}+\mathcal{A}^{(1)}+\mathcal{A}^{(2)}+\ldots$ in terms of the Dyson series representation of $\mathcal{U}$, so that  $\mathcal{A}^{(n)}\equiv\bra{\phi_\text{out};\psi_\text{out}}\mathcal{U}^{(n)}\ket{\phi_\text{in};\psi_\text{in}}$. Using
\begin{align}
\mathcal{U}^{(n)}=(-i)^n&\int^{}_{}\mathrm{d}t_1\int_{}^{t_1}\mathrm{d}t_{2}\ldots  \notag \\ \times&\int^{t_{n-1}}\mathrm{d}t_n   H_\text{I}(t_1)H_\text{I}(t_2)\ldots H_\text{I}(t_n),
\end{align}
we determine that the $n$\textsuperscript{th} order term in the transition amplitude contains $n$ copies of the interaction Hamiltonian. This will be an important observation in Secs.~\ref{sec:tls} and \ref{sec:lambda}, where the interaction Hamiltonian is of Jaynes-Cummings form and conserves excitation number \cite{1443594}. Here and throughout this article we adopt the convention that unspecified upper and lower integration limits correspond to $\infty$ and $-\infty$ respectively. 

\section{The Two Level System}
\label{sec:tls}
\noindent
In this section we explicitly calculate the transition amplitude $\mathcal{A}$ for the scenario where the local system is a single TLS with states $\{\ket{g},\ket{e}\}$ that are separated by the transition frequency $\Omega$ and coupled to bosonic modes of all frequencies equally at a rate $\gamma$. In App. \ref{sec:ham} we show that the interaction Hamiltonian for this system is given by ($\hbar=1$)
\begin{align}
H_\text{I}(t)=\gamma\int\mathrm{d}\epsilon \ (e^{-i\Delta_\epsilon t}\sigma_+a_\epsilon+e^{i\Delta_\epsilon t}\sigma_- a_\epsilon^\dagger), \label{eq:finham}
\end{align}
where the waveguide's central frequency (around which we linearise the dispersion relation) is denoted by $\omega_0$ and we define the detuning $\Delta_\epsilon\equiv\omega_0+\epsilon-\Omega$. We assume that the TLS is prepared in the ground state $\ket{g}$ and, as $t_f\rightarrow\infty$, it is also true that $\ket{\phi_\text{out}}=\ket{g}$. 

The form of Hamiltonian \eqref{eq:finham} and our assumption of an initially and finally relaxed emitter means that the only non-zero contributions to $\mathcal{A}^{(n)}$ are those where $n$ is even and the Pauli matrices are ordered as $\sigma_-\sigma_+\ldots\sigma_-\sigma_+$. The general expression for the $n$\textsuperscript{th} order term in the transition amplitude is then
\begin{align}
\mathcal{A}^{(n)}=(-i\gamma)^n\int&\mathrm{d}\tilde{t}^{(n)}\int\mathrm{d}\bar{\epsilon}^{(n)} \  e^{i(\Delta_{\epsilon_1}t_1- \Delta_{\epsilon_2}t_2+\ldots-\Delta_{\epsilon_n}t_n)} \notag \\ & \ \ \ \  \times \bra{\psi_\text{out}}a_{\epsilon_1}^\dagger a_{\epsilon_2}a^\dagger_{\epsilon_3}\ldots a_{\epsilon_n}\ket{\psi_\text{in}}, \label{eq:transamp}
\end{align}
where ${\int\mathrm{d}\tilde{t}^{(n)}\equiv\int\mathrm{d}t_1\int_{}^{t_1}\mathrm{d}t_2\ldots\int_{}^{t_{n-1}}\mathrm{d}t_n}$ and $\int\mathrm{d}\bar{\epsilon}^{(n)}\equiv\int\mathrm{d}\epsilon_1\int\mathrm{d}\epsilon_2\ldots\int\mathrm{d}\epsilon_n$.

\subsection{Single Photon Scattering}
\label{sec:tlsa}
\begin{figure}[t!]
	\begin{center}
		\includegraphics[scale=1]{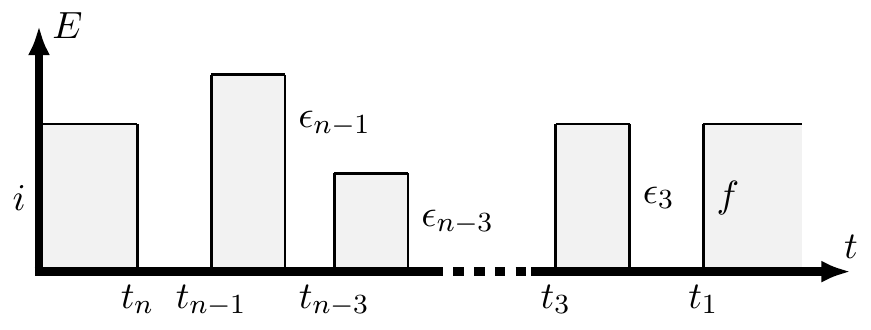}
		\caption{Diagram for the $n$\textsuperscript{th} order single photon scattering process. An incident photon of energy $\omega_0+i$ is scattered to one of frequency $\omega_0+f$. This occurs via the emission and absorption of $\frac{n}{2}-1$ `internal' photons.}
		\label{fig:spdiag}
	\end{center}
\end{figure}
\begin{figure}[t!]
	\begin{center}
		\includegraphics[width=7.5cm]{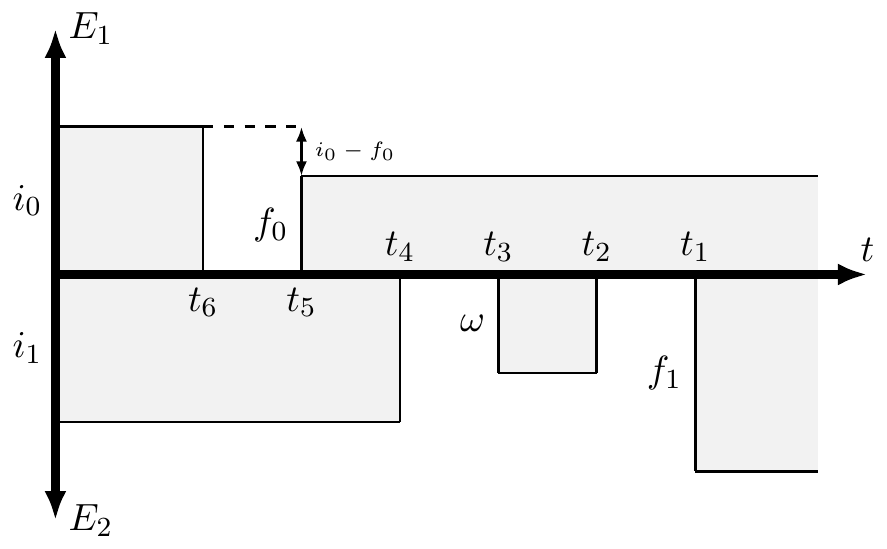}
		\caption{Diagram for one of the possible $n=6$ processes. Incident photons of energy $\omega_0+i_{0/1}$ are scattered to energies $\omega_0+f_{0/1}$. An internal photon `loop' of energy $\omega$ occurs, and $\omega$ is integrated over.
			\label{fig:drep}}
	\end{center}
\end{figure}
\noindent
We now demonstrate how to calculate the transition amplitude in Eq.~\eqref{eq:transamp} for the situation where a single incident photon with energy $\omega_0+i$ scatters to an output photon of energy $\omega_0+f$. It is simply a matter of applying the bosonic commutation relation to determine $\mathcal{A}^{(0)}=\delta(f-i)$. Consider now the $n$\textsuperscript{th} order term given by
\begin{align}
\mathcal{A}^{(n)}=\left(-i\gamma\right)^n\int\mathrm{d}\tilde{t}^{(n)}\int&\mathrm{d}\bar{\epsilon}^{(n)}\; e^{ i\Delta_{\epsilon_1}t_1 -i\Delta_{\epsilon_2}t_2\ldots-i\Delta_{\epsilon_n}t_n} \notag \\ \times&\bra{0}a_fa^\dagger_{\epsilon_1}a_{\epsilon_2}\ldots a_{\epsilon_n}a^\dagger_i\ket{0}, \label{eq:gentransamp}
\end{align}
we can use the vacuum expectation value in Eq.~\eqref{eq:gentransamp} to eliminate the first, final and half of the remaining frequency integrals
\begin{align}
\mathcal{A}^{(n)}=\left(-i\gamma\right)^n\int&\mathrm{d}\tilde{t}^{(n)}\int\mathrm{d}\epsilon_3 \mathrm{d}\epsilon_5\ldots\mathrm{d}\epsilon_{n-1}  \notag \\ \times &e^{i\left(\Delta_{f}t_1-\Delta_{\epsilon_3}t_2+\Delta_{\epsilon_3}t_3+\ldots\Delta_it_n\right)} \label{eq:formid}.
\end{align}
The integrand in \eqref{eq:formid} can be further decomposed into its constituent Dirac delta functions and we have then
\begin{align}
\mathcal{A}^{(n)}=\left(-i\gamma\right)^n(2\pi)^{\left(\frac{n}{2}-1\right)}\int\mathrm{d}t_1 \ e^{i\Delta_ft_1} \int_{}^{t_1}\mathrm{d}t_2 \notag \\ \times \int_{}^{t_2}\mathrm{d}t_3 \ \delta(t_3-t_2) \ldots \int_{}^{t_{n-1}}\mathrm{d}t_n \ e^{-i\Delta_it_n}.
\end{align}
Successively performing time integrals using the technique found in e.g. Ref.~\cite{Branczykphdthesis} and reproduced here in App.~\ref{sec:int} we arrive upon
\begin{align}
\mathcal{A}^{(n)}=2\left(-i\gamma\right)^n\delta(f-i)\left(\pi g(\Delta_i)\right)^\frac{n}{2},
\end{align} 
where we defined $g(\Delta)\equiv\left[\pi\delta(\Delta)+i\Delta^{-1}\right]$ for brevity. Summing over even $n$ and using the binomial theorem, we find
\begin{align}
\mathcal{A}=\frac{1-\gamma^2\pi g(\Delta_i)}{1+\gamma^2\pi g(\Delta_i)} \ \delta(f-i)\equiv t(i) \ \delta(f-i), \label{eq:finalsp}
\end{align}
 Eq.~\eqref{eq:finalsp} is valid under the condition $\protect{\left|\pi\gamma^2g(\Delta_i)\right|<1}$, which is required for application of the binomial theorem. However in App.~\ref{sec:Bor} we further use a Borel summation technique \cite{Suslov2005} to demonstrate the validity of the result for arbitrary values of $\protect{\left|\pi\gamma^2g(\Delta_i)\right|}$. Eq.~\eqref{eq:finalsp} is the first key result of this work and demonstrates that our method yields analytic expressions for the single-photon transition amplitude. In App. \ref{sec:Fan} we demonstrate its equivalence to the result of Fan \textit{et al} \cite{PhysRevA.82.063821}.

Note that it is quite easy to understand the nature of the physical process described by Eq.~\eqref{eq:formid} and we have sketched it explicitly in Fig. \ref{fig:spdiag}. We see that the atom absorbs the original incident photon and, before emitting the outgoing photon, emits and absorbs $\frac{n}{2}-1$ photons of frequencies $\left\{\epsilon_{n-1},\epsilon_{n-3}\ldots\epsilon_3\right\}$. The energies of these `internal' photons are uncertain and we integrate over a continuum of possible values for each, which has the effect of reducing their duration to zero---a `point-like' interaction.

\subsection{Two Photon Scattering}
\begin{figure}[t!]
	\begin{center}
		\includegraphics[width=7.5cm]{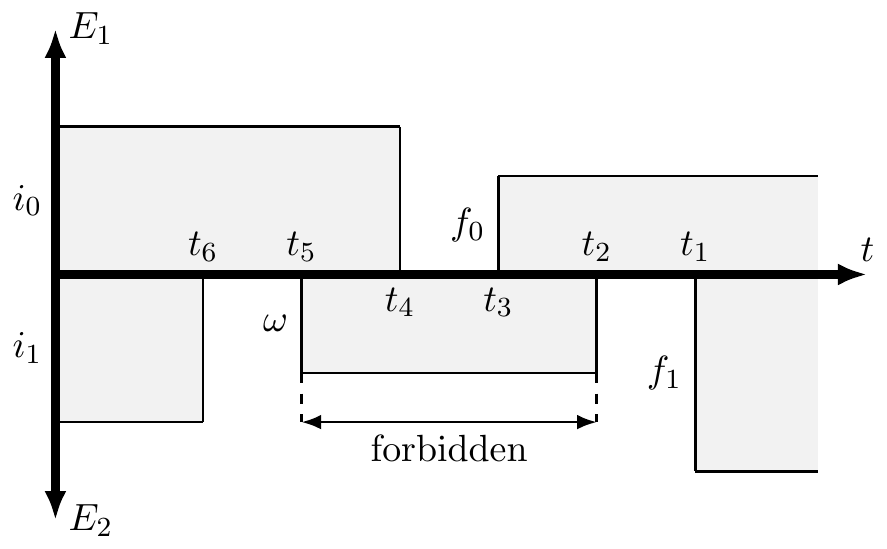}
		\caption{Diagram for one of the impossible $n=6$ processes. Incident photons of energy $\omega_0+i_{0/1}$ are scattered to energies $\omega_0+f_{0/1}$. An internal photon `loop' of energy $\omega$ occurs, and $\omega$ is integrated over.
			\label{fig:forbid}}
	\end{center}
\end{figure}
\label{sec:tp}
\noindent
In this section we elaborate further on the diagrammatic method alluded to in Sec. \ref{sec:tlsa} in order to evaluate the two-photon transition amplitude. For input photons with energies $\omega_0+i_0$ and $\omega_0+i_1$, the $n$\textsuperscript{th} order term in the transition amplitude is
\begin{align}
\mathcal{A}^{(n)}=\left(-i\gamma\right)^n\int&\mathrm{d}\tilde{t}^{(n)}\int\mathrm{d}\bar{\epsilon}^{(n)}  \ e^{i(\Delta_{\epsilon_1}t_1-\Delta_{\epsilon_2}t_2\ldots-\Delta_{\epsilon_n}t_n)} \notag \\ \times &\bra{0}a_{f_0}a_{f_1}a^\dagger_{\epsilon_1}a_{\epsilon_2}\ldots a_{\epsilon_n}a^\dagger_{i_1}a^\dagger_{i_0}\ket{0}, \label{eq:gentransamp2}
\end{align}
where $f_0$ and $f_1$ label the scattered photon frequencies. Evaluation of the vacuum expectation value in the integrand of Eq.~\eqref{eq:gentransamp2} produces $2^{\frac{n}{2}+1}$ terms \cite{PhysRev.80.268} and it is not feasible to mechanically calculate these. We instead use the physical interpretation of each term to provide further guidance. 

\begin{figure*}[t!]
	\subfloat[]{
		\includegraphics[scale=0.65]{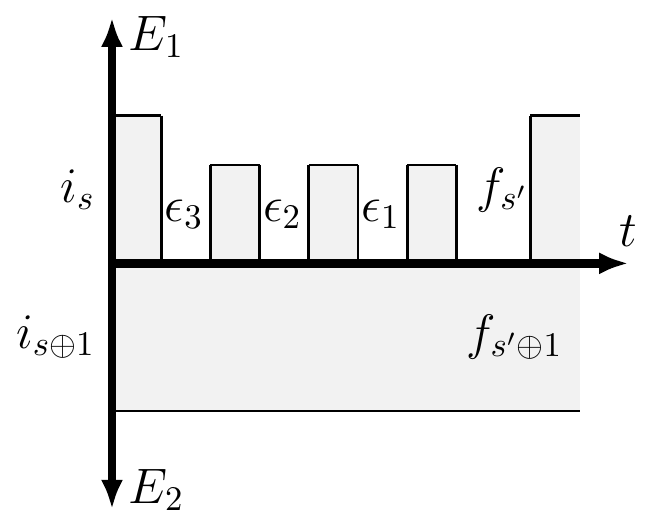}\label{fig:nomix}}
	\subfloat[]{
		\includegraphics[scale=0.65]{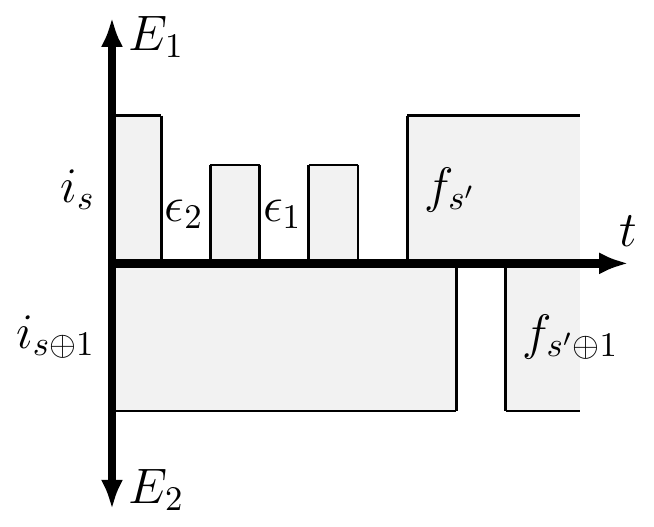}\label{fig:b}}
	\subfloat[]{
		\includegraphics[scale=0.65]{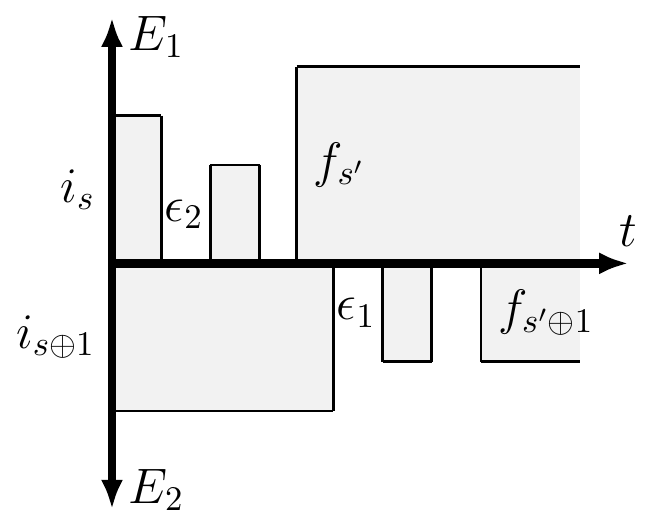}\label{fig:c}}
	\subfloat[]{
		\includegraphics[scale=0.65]{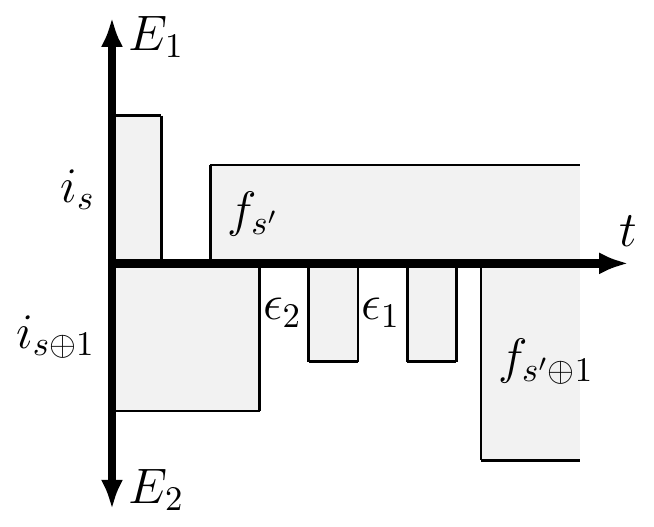}\label{fig:d}}
	\caption{The four non-zero types of diagram for the $\mathcal{A}^{(8)}$ term in the expansion of the two photon amplitude. Fig. \ref{fig:nomix} represents the non-frequency mixing term.}
	\label{fig:diagrams}
\end{figure*}

For example, consider one of the sixteen terms contributing to $\mathcal{A}^{(6)}$
\begin{align}
\mathcal{A}^{(6)}_{(1)}=-&\gamma^6\int\mathrm{d}\tilde{t}^{(6)}  \int\mathrm{d}\omega \notag \\ \times &e^{i(\Delta_{f_1}t_1-\Delta_\omega (t_2-t_3)-\Delta_{i_1}t_4+\Delta_{f_0} t_5-\Delta_{i_0}t_6)},
\label{eq:int}
\end{align}
which, using exactly the same integration techniques as for the single-photon case, reduces to
\begin{align}
\mathcal{A}^{(6)}_{(1)}=-2\pi^2\gamma^6&\delta(f_0+f_1-i_0-i_1) \notag \\\times g(\Delta_{i_0})  &g(\Delta_{i_0}-\Delta_{f_0})g(\Delta_{i_0}+\Delta_{i_1}-\Delta_{f_0})^2.
\label{eq:reduc}
\end{align}
By re-associating bosonic mode operators to their phases in the integrand of Eq.~\eqref{eq:int} we deduce that this term describes absorption by the atom of a photon with energy $\omega_0+i_0$, prior to emission of a final $\omega_0+f_0$ photon. Subsequently, the second incident photon is absorbed and emitted twice via an intermediate step of energy $\omega_0+\omega$. Fig.~\ref{fig:drep} gives a pictorial representation of the process, with time evolving from left-to-right and energies of the two populated modes relative to $\omega_0$ given by the distance from the horizontal axis.

We can derive amplitudes in general from diagrams such as Fig.~\ref{fig:drep}. By drawing the diagrams corresponding to the possible emission/absorption processes we can calculate the total transition amplitude.  With each emission and absorption event in a diagram we associate a number $\Delta$ representing the difference between the total amount of absorbed radiation by the atom and the ground-excited energy gap. In Fig.~\ref{fig:drep}, the atom absorbs a photon of frequency $\omega_0+i_0$ (yielding $\Delta_{i_0}$), and emits a photon with energy $\omega_0+f_0$ yielding $\Delta_{i_0}-\Delta_{f_0}$ corresponding to the residual energy between the two photons. Absorbing the second incident photon produces the factor $\Delta_{i_0}+\Delta_{i_1}-\Delta_{f_0}$. These terms appear as arguments of the frequency dependent function $g(x)$ in Eq.~\eqref{eq:reduc}, which describes the amplitude of the process depicted in Fig.~\ref{fig:drep}. The `loop' indicated by $\omega$ in Fig.~\ref{fig:drep} increases the power of $g(\Delta_{i_0}+\Delta_{i_1}-\Delta_{f_0})$  by one. Finally, we impose energy conservation via $\delta(f_0+f_1-i_0-i_1)$.

Suppose that for a given $n$ we have drawn all diagrams corresponding to $\frac{n}{2}$ light-matter interaction events. Four of these diagrams (the permutations over initial and final photon frequencies) will always have one photon interacting with the emitter $\frac{n}{2}$ times, with the second photon passing through unperturbed (i.e., non-frequency mixing terms). These diagrams  contribute amplitudes equivalent to the single photon case. Another class of diagrams we immediately discard is that in which an `internal' photon (such as $\omega$ in Fig.~\ref{fig:drep}) is emitted at time $t_m$ and \emph{not} reabsorbed at $t=t_{m-1}$, since the interval $[t_m,t_{m-1}]\rightarrow0$. We rigorously demonstrate this in App. \ref{sec:zer}. The remaining diagrams are similar in structure to Fig.~\ref{fig:drep}, with initial absorption and final emission separated by a number of internal photon loops. The structure of the integrals corresponding to these diagrams is the same as in Eq.~\eqref{eq:int} with additional frequency and time integrals corresponding to these internal loops.

The procedure for converting diagrams into $\mathcal{A}^{(n)}$ is as follows:
\begin{itemize}\itemsep=0pt
	\item[(i)] draw all possible diagrams with $\frac{n}{2}$ total interactions; 
	\item[(ii)] identify the single photon (non-frequency mixing) terms; 
	\item[(iii)] discard the terms in which internal photons are emitted  and not immediately reabsorbed; 
	\item[(iv)] the remaining terms get the constant pre-factor $\frac{2}{\pi}(i\sqrt{\pi}\gamma)^n$; 
	\item[(v)] each absorption event gets a factor $g(\Delta)$, where $\Delta$ corresponds to the total absorbed radiation, and each emission event gets $g(\Delta_\text{res})$, where $\Delta_{\text{res}}$ is the amount of absorbed radiation not re-emitted;
	\item[(vi)] for each loop, multiply by an additional factor of $g(\Delta)$ with the same $\Delta$ as at the previous absorption; 
	\item[(vii)] at the final emission, multiply by $\delta(f_0+f_1-i_0-i_1)$.
\end{itemize}
The four species of diagram for the $n=8$ case are shown in Fig. \ref{fig:diagrams} and in App. \ref{sec:exp} we explicitly perform this procedure to demonstrate equivalence between the diagrammatic and integral methods. 

One interesting observation here is that for $n\geq6$ the particular form of Eq.~\eqref{eq:finham} causes vanishing of the terms with internal photon emission not immediately followed by re-absorption (step (iii) of the above outlined rules). This behaviour is due to the Hamiltonian's instantaneous coupling between the emitter and continuum of {waveguide} modes (without cut-off) at a constant rate. It is interesting to note that this oft-employed model makes this prediction and still agrees well with experimental data. General Hamiltonians with discretised waveguide modes would not necessarily lead to these terms vanishing. We show one of these dis-allowed diagrams in Fig. \ref{fig:forbid}.

Let the frequency mixing term in $\mathcal{A}^{(n)}$ for the two-photon case be given by $\delta(f_0+f_1-i_0-i_1)\mathcal{M}^{(n)}$. From the above procedure we deduce that the total photon frequency mixing term in the two-photon transition amplitude is given by $\mathcal{M}=\sum_{n=2}^{\infty}\mathcal{M}^{(n)}$ where:
\begin{align}
\mathcal{M}^{(n)}=&\sum_{s=0,1}^{}\sum_{s'=0,1}^{}g(\Delta_{i_s})g(\Delta_{i_s}-\Delta_{f_{s'}})g(\Delta_{f_{s'\oplus1}}) \notag \\ \qquad \ \ \ \  \times&\frac{2}{\pi}(-\pi\gamma^2)^n\sum_{k=0}^{n-2}g(\Delta_{i_s})^kg(\Delta_{f_{s'\oplus1}})^{n-2-k}
\label{eq:withsum}
\end{align}
The sum over $k$ can be evaluated \cite{spivak1980calculus}, and we sum over all $n$ to find $\mathcal{M}$. Adding this to the non-frequency mixing component yields a final expression for the two-photon transition amplitude:
\begin{align}
\mathcal{A}=&[t(i_0)+t(i_1)-1]\notag\\&\times[\delta(f_0-i_0)\delta(f_1-i_1)+\delta(f_0-i_1)\delta(f_1-i_0)] \notag\\ +&2\pi\gamma^4\delta(f_0+f_1-i_0-i_1)\sum_{s=0,1}^{}\sum_{s'=0,1}\notag\\\times&  \frac{g(\Delta_{i_s})g(\Delta_{i_s}-\Delta_{f_{s'}})g(\Delta_{f_{s'\oplus1}})}{[1+\pi\gamma^2g(\Delta_{i_s})][1+\pi\gamma^2g(\Delta_{f_{s'\oplus1}})]}, \label{eq:fintp}
\end{align}
where $t(i)$ is defined in Eq.~\eqref{eq:finalsp}. Eq.~\eqref{eq:fintp} is the second main result of this work and demonstrates our formalism's power to produce non-perturbative amplitudes for multi-photon processes. We demonstrate its equivalence the expression found by Fan \textit{et al} in App. \ref{sec:Fan}.

\section{$\Lambda$-System}
\label{sec:lambda}
\begin{figure}[t!]
	\begin{center}
		\includegraphics[scale=1]{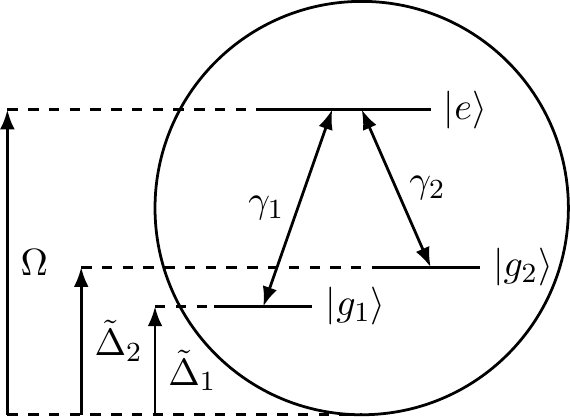}
	\end{center}
	\caption{A so-called $\Lambda$-system. Two ground levels $\ket{g_1}$ and $\ket{g_2}$ are coupled with amplitudes $\gamma_1$ and $\gamma_2$ respectively to an excited state $\ket{e}$. We define a zero-energy separated from $\ket{e}$ by $\Omega$ and denote the gap between $\ket{g_1}$/$\ket{g_2}$ and zero by $\tilde{\Delta}_1$/$\tilde{\Delta}_2$. In the following we assume $\Omega>\tilde{\Delta}_2>\tilde{\Delta}_1$.}
	\label{fig:lambda}
\end{figure}
\noindent
In many cases the perfect TLS is hard to realise, or some additional control is required over the system. This means that the emitter used in many light-matter interaction experiments has a more complex internal structure, e.g. in Refs.~\cite{Shomroni903,Reiserer1349,Dawes672}. This motivates the extension of our method to a second species of local system. Consider the $\Lambda$-system shown schematically in Fig.~\ref{fig:lambda}. Neglecting polarisation, the interaction Hamiltonian describing the dynamics of this system is readily derived \cite{PhysRevA.95.063809} and given by
\begin{align}
H_\text{I}(t)=\sum_{\lambda=1}^{2}\gamma_\lambda\int\mathrm{d}\epsilon \ e^{it\Delta_{\epsilon,\lambda}}a_\epsilon^\dagger\ket{g_\lambda}\!\bra{e}+e^{-it\Delta_{\epsilon,\lambda}}a_\epsilon\ket{e}\!\bra{g_\lambda} \label{eq:Ham}
\end{align}
where we have defined the detuning $\Delta_{\epsilon,\lambda}=\omega_0+\epsilon-\Omega+\tilde{\Delta}_\lambda$, again linearising the waveguide dispersion relation about $\omega_0$. In general, prior to and following a photon scattering event, a $\Lambda$-system will be in some state described by $\ket{\phi}=\alpha\ket{g_1}+\beta\ket{g_2}$, as radiative transitions to each of the ground states are allowed but the $\ket{g_1}\leftrightarrow\ket{g_2}$ transitions are forbidden. In order to fully specify the dynamics then, we need to evaluate matrix elements of the form
\begin{align}
\mathcal{A}_{\mu\nu}=\bra{\psi_\text{out};g_\mu}\mathcal{U}\ket{\psi_\text{in};g_\nu},
\end{align}
where $\mu/\nu=1,2$.
Inserting Hamiltonian \eqref{eq:Ham} into this expression for the transition amplitude then yields
\begin{align}
\mathcal{A}_{\mu\nu}^{(n)}=(-i)^n\sum_{\{\lambda_1,\lambda_2\ldots\lambda_n\}=1}^{2}\gamma_{\lambda_1}\gamma_{\lambda_2}\ldots\gamma_{\lambda_n}  \int\mathrm{d}\tilde{t}^{(n)}\int\mathrm{d}\bar{\epsilon}^{(n)} \notag \\ \bra{\psi_\text{out};g_\mu}  e^{it_1\Delta_{\epsilon_1,\lambda_1}}a_{\epsilon_1}^\dagger   \ket{g_{\lambda_1}}\!\bra{e}e^{-it_2\Delta_{\epsilon_2,\lambda_2}}a_{\epsilon_2}\ket{e}\!\bra{g_{\lambda_2}}\ldots \notag \\  \times e^{-it_n\Delta_{\epsilon_n,\lambda_n}}a_{\epsilon_n}\ket{e}\!\bra{g_{\lambda_n}}\ket{\psi_\text{in};g_\nu}, \label{eq:amp}
\end{align}
where, at each time step, we inserted only the two terms from the Hamiltonian which either raise a ground state or lower an excited one---the two terms corresponding to the opposite behaviour necessarily vanishing. This means that again, Eq.~\eqref{eq:amp} is non-zero only when $n$ is even. The final simplification Eq.~\eqref{eq:amp} permits, before requiring knowledge about the input and output optical states, utilises the orthogonality of atomic states to eliminate $\frac{n}{2}+1$ of the sums over $\lambda$ by replacing inner products between ground states with Kronecker Delta functions, e.g. $\braket{g_{\lambda_2}|g_{\lambda_3}}=\delta_{\lambda_2\lambda_3}$.

\subsection{Single Photon Scattering}
\noindent
We can evaluate the amplitude of Eq.~\eqref{eq:amp} for the case of single photon scattering. We denote the input and output optical states by $\ket{\psi_\text{in}}=\ket{i}$ and $\ket{\psi_\text{out}}=\ket{f}$ respectively. It is simple to deduce that 
\begin{align}
&\bra{\psi_\text{out}}a_{\epsilon_1}^\dagger a_{\epsilon_2}\ldots a_{\epsilon_n}\ket{\psi_\text{in}}=\bra{0}a_fa_{\epsilon_1}^\dagger a_{\epsilon_2}\ldots a_{\epsilon_n}a_i^\dagger\ket{0} \notag \\ =&\delta(f-\epsilon_1)\delta(\epsilon_2-\epsilon_3)\ldots\delta(\epsilon_{n-2}-\epsilon_{n-1})\delta(\epsilon_n-i)
\end{align}
and we can therefore eliminate $\frac{n}{2}+1$ of the integrals over $\epsilon$ in Eq.~\eqref{eq:amp}, leaving
\begin{align}
\mathcal{A}^{(n)}_{\mu\nu}=(-i)^n\gamma_\mu\gamma_\nu\sum_{\{\lambda_2,\lambda_4\ldots\lambda_{n-2}\}=1}^{2}\gamma_{\lambda_2}^2\gamma_{\lambda_4}^2\ldots\gamma_{\lambda_{n-2}}^2\int\mathrm{d}\tilde{t}^{(n)} \notag \\ \times \int\mathrm{d}\epsilon_2\int\mathrm{d}\epsilon_4\ldots\int\mathrm{d}\epsilon_{n-2}  e^{it_1\Delta_{f,\mu}}e^{-it_2\Delta_{\epsilon_2,\lambda_2}}e^{it_3\Delta_{\epsilon_2,\lambda_2}}\ldots \notag \\ \times e^{it_{n-1}\Delta_{\epsilon_{n-2},\lambda_{n-2}}}e^{-it_n\Delta_{i,\nu}}. \label{eq:part}
\end{align}
Successively evaluating the frequency integrals in Eq.~\eqref{eq:part} in the same manner as for the TLS case, we find
\begin{align}
\mathcal{A}^{(n)}_{\mu\nu}=2(-i)^n&[\pi(\gamma_1^2+\gamma_2^2) g(\Delta_{i,\nu})]^{\frac{n}{2}}\frac{\gamma_\mu\gamma_\nu}{\gamma_1^2+\gamma_2^2}\delta(\Delta_{f,\mu}-\Delta_{i,\nu})
\end{align}
and again apply the binomial theorem/Borel summation to determine
\begin{align}
\mathcal{A}_{\mu\nu}&=\delta(\Delta_{f,\mu}-\Delta_{i,\nu})\left(\delta_{\mu\nu}-\frac{2i\pi\gamma_\mu\gamma_\nu}{\Delta_{i,\nu}+i\pi(\gamma_1^2+\gamma_2^2)}\right) \notag \\ &\equiv \delta(\Delta_{f,\mu}-\Delta_{i,\nu})\left[\delta_{\mu\nu}+s_{\mu\nu}(\Delta_{i\nu})\right] . \label{eq:SPLS}
\end{align}

The predictions of Eq.~\eqref{eq:SPLS} can be arrived upon via a variety of other methods, e.g. Refs.~\cite{PhysRevA.80.033823,1367-2630-12-4-043052,PhysRevA.85.043814,1367-2630-15-11-115010}. Specifically we see that Eqs.~(23) of Ref.~\cite{PhysRevA.85.043814} are recovered under the transformation $\pi\gamma_i^2\rightarrow\Gamma_i$. For a $\Lambda$-system with identical lifetimes into both ground states, i.e. $\gamma_1=\gamma_2$, the prediction that a single resonant photon, incident upon an emitter prepared in the state $\ket{g_1}$ deterministically transfers the population to the state $\ket{g_2}$ is reproduced. 

\subsection{Two Photon Scattering}
\noindent
We now argue that is possible to extend the diagrammatic approach used to compute the two-photon transition amplitude for the TLS to the $\Lambda$-system. In order to do this we need to demonstrate that the rules enumerated in Sec.~\ref{sec:tp} continue to apply---with slight modifications specified by the added internal structure of the emitter. The first task therefore is to show that we can continue to discard terms in which internal photons are emitted and not immediately reabsorbed. These diagrams correspond to terms in the transition amplitude where an integral over the continuum of modes leads to a delta function connecting non-adjacent times. It is easy to determine that this continues to be the case by inspection of Eq.~\eqref{eq:amp}. We see that the structure of the time integral is not modified and so any delta function in the integrand of the form $\delta(t_i-t_j)$, where $|i-j|>1$, will continue to integrate to zero by the logic of App.~\ref{sec:zer}. 

The non-frequency mixing diagrams for the $\Lambda$-system are again simple to analyse but yield a subtly different term to that found in the TLS case. This is expected \cite{PhysRevA.95.063809} and related to the breaking of photon exchange symmetry, introduced by the non-unique ground states of the atomic system. Non-frequency mixing diagrams correspond to the four terms in the transition amplitude where, when the vacuum expectation value in Eq.~\eqref{eq:amp} is evaluated, one of the creation operators for an initial photon state is commuted through one of the operators for a final state photon. This means that the structure of delta functions in the integrand of such terms is
\begin{align}
\delta(f'-i')\delta(f-\epsilon_1)\delta(\epsilon_2-\epsilon_3)\delta(\epsilon_4-\epsilon_5)\ldots\delta(\epsilon_n-i),
\end{align} 
where $f/f'$ and $i/i'$ label the frequencies of final and initial state photons respectively. We can therefore construct the frequency-preserving portion of the transition amplitude
\begin{align}
\mathcal{N}_{\mu\nu}=\delta_{\mu\nu}\left[\delta(f_0-i_0)\delta(f_1-i_1)+\delta(f_0-i_1)\delta(f_1-i_0)\right] \notag \\ +\sum_{s=0,1}\sum_{s'=0,1}\delta(f_{s'\oplus1}-i_{s\oplus1})\delta(\Delta_{f_{s'}\mu}-\Delta_{i_s\nu})s_{\mu\nu}(\Delta_{i_s\nu}).
\end{align}

\begin{figure*}
	\subfloat[TLS and special $\Lambda$-system configurations. Central angular frequencies are $\Omega=2\times10^{15}$, $2.2\times10^{15}$ and $2.4\times10^{15}$ rads$^{-1}$ for systems represented by circles, triangles and crosses respectively. The coupling $\gamma_1=2\times10^4$ (rad/s)$^\frac{1}{2}$. ]{\includegraphics[scale=1]{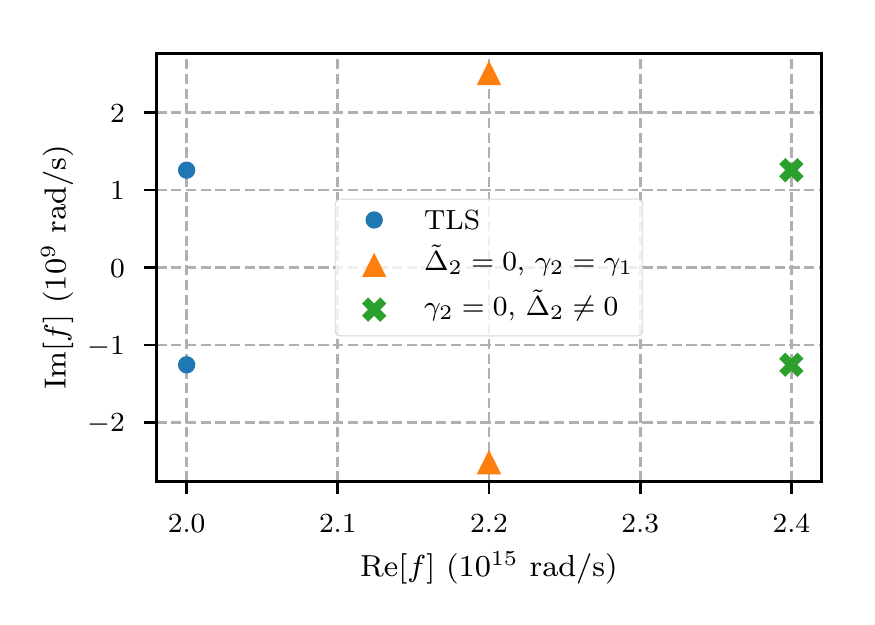} \label{fig:tlspoles}}
	\subfloat[$\Lambda$-system prepared initially in the state $\ket{g_1}$, with $\Omega=2\times10^{15}$ rad/s, $\tilde{\Delta}_1=0$ and $\tilde{\Delta}_2=\Omega$/10. We further set $\gamma_1=2\times10^4$ (rad/s)$^\frac{1}{2}$ and $\gamma_2=\gamma_1$/$\sqrt{2}$. Some poles correspond specifically to the emitter scattering to a given state and others are present in both cases.]{\includegraphics[scale=1]{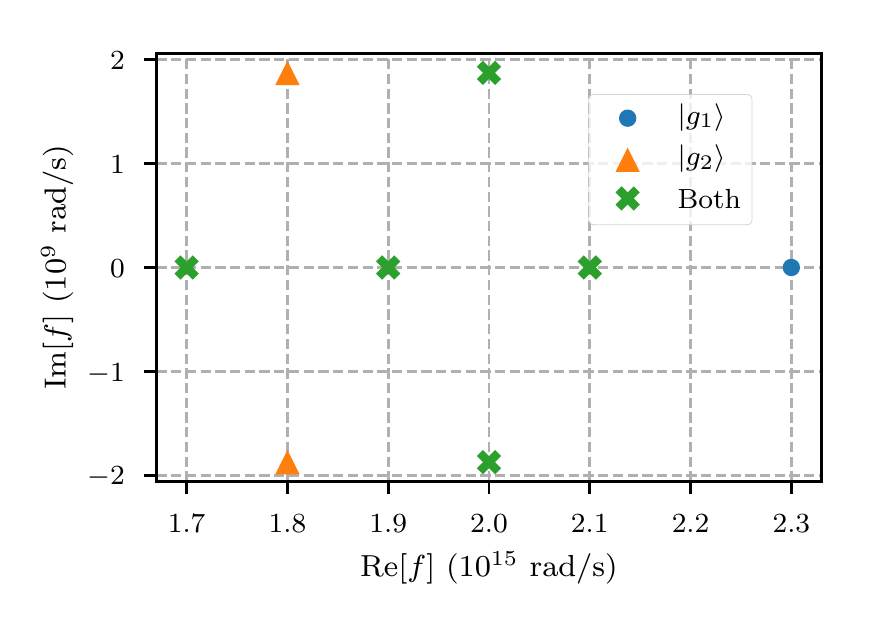} \label{fig:lpoles}}
	\caption{Location of poles in the photon mixing component of the total transition amplitude on the complex energy plane of $f$---the energy of one of the scattered photons. The coupling \protect{$\gamma_1=2\times10^4$ (rad/s)$^{\frac{1}{2}}$} corresponds to a lifetime of approximately $1$ ns. In both cases we drive the system with two single-frequency photons, one positively detuned from $\Omega$ by $\delta=1\times10^{14}$ rad/s and one negatively detuned by the same amount.}
	\label{fig:poles}
\end{figure*}

Having evaluated the non-frequency mixing terms and also those which do not contribute to the transition amplitude, the only species of terms remaining correspond to frequency mixing processes. Applying the constraint that internal photons must be immediately reabsorbed following emission, we find that the structure of the vacuum expectation value in the integrand of frequency mixing terms is
\begin{align}
\delta(\epsilon_n-i)\delta(\epsilon_{n-1}-\epsilon_{n-2})\ldots   \delta(f-\epsilon_{m+1})\delta(i'-\epsilon_m) \notag \\ \times \delta(\epsilon_{m-1}-\epsilon_{m-2})\ldots\delta(f'-\epsilon_1),
\end{align}
where $m$ labels some point along the time evolution where one photon ceases its interaction with the emitter and the second one is absorbed. This completes our argument, as we see that again in order to calculate the $n$\textsuperscript{th} order term in the transition amplitude we have to sum all terms with $\frac{n}{2}$ total interactions, varying the number of times each of the initial photons interacts. Performing this procedure we calculate the $n$\textsuperscript{th} order frequency mixing term
\begin{align}
\mathcal{M}_{\mu\nu}^{(n)}=&\frac{2}{\pi}(-\pi)^n\gamma_\mu\gamma_\nu(\gamma_1^2+\gamma_2^2)^{n-2}\sum_{s=0,1}^{}\sum_{s'=0,1}^{}\sum_{\lambda=1,2}^{}\gamma_\lambda^2 \notag \\ \times &g(\Delta_{i_s\nu})g(\Delta_{i_s\nu}-\Delta_{f_{s'}\lambda})g(\Delta_{f_{s'\oplus1}\mu})\notag \\ \times &\sum_{k=0}^{n-2}g(\Delta_{i_s\nu})^{n-2-m}g(\Delta_{f_{s'\oplus1}\mu})^m \notag \\ \times &\delta(\Delta_{f_{s'\oplus1}\mu}+\Delta_{f_{s'}}-\Delta_{i_{s\oplus1}}-\Delta_{i_s\nu}), \label{eq:nth}
\end{align}
which we see is similar in structure to Eq.~\eqref{eq:withsum} with an additional sum over the two possible mechanisms by which the two incident photons could now couple. After summing expression Eq.~\eqref{eq:nth} over all $n$, adding this to the frequency preserving term and algebraic rearrangement we find the total transition amplitude
\begin{align}
\mathcal{A}_{\mu\nu}=\mathcal{N}_{\mu\nu} +\frac{1}{2\gamma_\mu\gamma_\nu}\sum_{s,s',\lambda}^{}\gamma_\lambda^2s_{\mu\nu}(\Delta_{i_s\nu})s_{\mu\nu}(\Delta_{f_{s'\oplus1}\mu}) \notag \\ \times\delta(\Delta_{i_{s\oplus1}\lambda}-\Delta_{f_{s'\oplus1}\mu})\delta(\Delta_{i_s\nu}-\Delta_{f_{s'}\lambda}) \notag \\ +\frac{i}{2\pi\gamma_\mu\gamma_\nu}\sum_{s,s',\lambda}^{}\gamma_\lambda^2\frac{1}{\Delta_{i_s\nu}-\Delta_{f_{s'}\lambda}}s_{\mu\nu}(\Delta_{i_s\nu}) \notag \\ \times s_{\mu\nu}(\Delta_{f_{s'\oplus1}\mu})\delta(\Delta_{i_s\nu}-\Delta_{f_{s'\oplus1}\mu}+\Delta_{i_{s\oplus1}}-\Delta_{f_{s'}}). \label{eq:finaltpl}
\end{align}
The transition amplitude of Eq.~\eqref{eq:finaltpl} exactly specifies the combined emitter-optical state following the scattering of two initial photons with frequencies $i_0$ and $i_1$ on the $\Lambda$-system depicted in Fig.~\ref{fig:lambda}. We can use this to investigate the properties of light-matter scattering experiments and we do this in the following section. 

Fewer reported techniques exist that capture the physics of Eq.~\eqref{eq:finaltpl}, compared with the single-photon case. However, methods derived from those of relativistic quantum field theory do exist as in e.g. Ref.~\cite{1367-2630-14-9-095028}. Here Pletyukhov and Gritsev derive an expression for the `$T$-matrix,' $T^{(2)}(\omega)$ when two photons scatter from a $\Lambda$-system. In App.~\ref{app:njp} we demonstrate that Eq.~\eqref{eq:finaltpl} of this paper is equivalent to Eq.~(46) of Ref.~\cite{1367-2630-14-9-095028}. 

\section{Pole Structure of the Amplitude}
\label{sec:entangle}

\noindent
Consider a two-photon scattering experiment. It is known that the properties of the scattered state are determined in large part by the pole structure of the transition amplitude \cite{weinberg1995quantum}. In particular, poles in the complex plane of the scattered photon energy correspond to bound-states of the system \cite{taylor2012scattering}. We might naively imagine that the pole structure of the amplitude is broadly similar whether the two-photons scatter from a TLS or a $\Lambda$-system, with the added internal structure of the emitter only slightly shifting their location for example. We can however demonstrate that this is not the case and that the addition of a second emitter ground-state introduces a great deal of richness to the system. In Fig.~\ref{fig:poles} we consider the frequency-mixing portion of the transition amplitude of Eq.~\eqref{eq:finaltpl} and plot poles in the complex plane of $f$, which gives the energy of one of the scattered photons. Note that given $f$, the energy of the second photon is completely specified by the single energy conserving delta function. In both Figs.~\ref{fig:tlspoles} and \ref{fig:lpoles} we drive the system with two single-frequency photons, one detuned negatively from the transition energy $\Omega$ by $\delta=1\times10^{14}$ rad/s and one positively by the same amount. 

In Fig.~\ref{fig:tlspoles} we plot the location of the poles for three different systems. Blue circles illustrate the locations of the poles for a simple TLS, with central frequency $\Omega=2\times10^{15}$ rad/s and coupling $\gamma=2\times10^{4}$ (rad/s)$^{\frac{1}{2}}$. This coupling strength would correspond to a lifetime of 1 ns, which is a reasonable estimate for a TLS formed by e.g. a semiconductor quantum dot. We note the two poles at $f=\Omega\pm i\pi\gamma^2$, this is the result found by many previous authors and corresponds to the formation of a frequency-entangled pair of photons. It is interesting to ask under which circumstances the photons scattered from a $\Lambda$-system appear indistinguishable from those scattered by a TLS. Obviously, we would expect that when $\gamma_2=0$, for arbitrary $\tilde{\Delta}_2$, the system should behave as the TLS---photons have no access to the state $\ket{g_2}$. As a validity check of Eq.~\eqref{eq:finaltpl} we plot the poles of such as a system (with $\Omega=2.4\times10^{15}$ rad/s now) using green crosses and find that this is indeed the predicted behaviour. A more surprising result is indicated by the orange triangles of Fig.~\ref{fig:tlspoles}. Here we set $\gamma_2=\gamma_1$, with $\gamma_1$ the same as for the TLS. We find that, when $\tilde{\Delta}_2=0$, the pole structure of the $\Lambda$-system is again the same as that of the TLS---though the poles are now located at $f=\Omega\pm i\pi(\gamma_1^2+\gamma_2^2)$. This is due to the degenerate ground states appearing indistinguishable to incoming photons and thus their only effect is a strengthening of the light-matter interaction, evidenced by shifting of the poles away from the real axis. 

It is not generally true that the dynamics of scattering from a $\Lambda$-system are well approximated by the TLS. In Fig.~\ref{fig:lpoles} we consider a more general $\Lambda$-system with $\Omega=2\times10^{15}$ rad/s, $\tilde{\Delta}_1=0$ and $\tilde{\Delta}_2=\Omega/10$. We further assume the system is prepared initially in the lower ground state $\ket{g_1}$ and set the couplings asymmetrically so that again $\gamma_1=2\times10^{4}$ (rad/s)$^\frac{1}{2}$ but $\gamma_2=\gamma_1/\sqrt{2}$. Now, it is important to note that the frequency mixing component of Eq.~\eqref{eq:finaltpl} corresponds to two distinct processes. In one the emitter returns to the state $\ket{g_1}$ following the scattering, while in the other it scatters to $\ket{g_2}$. We plot both species of poles in Fig.~\ref{fig:lpoles}, using blue circles and orange triangles respectively and also use green crosses to denote the location of poles common to both parts of the transition amplitude. 

The most striking feature of Fig.~\ref{fig:lpoles} compared to \ref{fig:tlspoles} is the emergence of poles on the Im$[f]=0$ axis of the complex plane. This means that there are now singularities in the transition amplitude corresponding to physical scattered photon energies---resonances. These occur at the frequencies of the photons input to the system, plus the input frequencies minus the energy gap $\tilde{\Delta}_2$. For the emitter state preserving portion of the amplitude there is an additional resonance at $\Omega+\tilde{\Delta}_2+\delta$, stemming from a process where one of the photons scatters the system to $\ket{g_2}$, with the second photon then picking up this excess energy. The final point to note is the emergence of a second pair of imaginary poles at $\Omega-\tilde{\Delta}_2\pm i\pi\left(\gamma_1^2+\gamma_2^2\right)$ in the portion of the transition amplitude in which the emitter is scattered to the state $\ket{g_2}$. This has a simple physical interpretation; if we were to post-select onto this state, the bound state of entangled photons that formed would have its central frequency shifted so as to conserve overall energy. 

\section{Conclusions}
\label{sec:conc}
\noindent
We have developed an intuitive, diagrammatic approach to the problem of light-matter coupling in waveguide QED. In contrast to previously reported techniques, our method allows visualisation of the photon-atom dynamics. We have demonstrated analytical results for both single and two photon input optical states for both the TLS and $\Lambda$-systems. The diagrammatic approach is straight-forward to extend to higher photon number input states (though increasingly computationally expensive) and potentially more realistic Hamiltonians, and analytic results are expected to follow. Several open questions emerge from this work. For instance, how does the choice of Hamiltonian in Eq.~\eqref{eq:finham} impact the transition amplitude? In particular, a waveguide will have a range of supported frequency modes defined largely by its dimensions. In theory this leads to observable consequences \cite{PhysRevA.88.013836} and this would seem to suggest that some of the processes associated with forbidden diagrams actually contribute in physical systems. The limit on our method is ultimately a computational one, with an $N$-photon event requiring $N$ permutations over both initial and final frequencies. 

\section*{Acknowledgements}
\noindent
DLH acknowledges useful suggestions from M.E.~Pearce and G.~Ferenczi and is supported by an EPSRC studentship.

\appendix

\section{The Hamiltonian}
\label{sec:ham}
\noindent
In this appendix we derive the interaction Hamiltonian \eqref{eq:finham} that describes a TLS coupled to an optical waveguide. We begin by dividing the total Hamiltonian into free and interacting parts, $H_0$ and $H_\text{int}$ respectively. The dynamics of an isolated emitter and bare waveguide are described by $H_0$, while the coupling between them---which we assume is of dipole form---is specified by $H_\text{int}$. We take a limit where the waveguide supports a continuum of optical modes with wavenumber $k$ and apply the rotating-wave-approximation. This leads to
\begin{align}
{H}_0&=\frac{1}{2}\Omega\sigma_z+\int_{0}^{}\mathrm{d}k \ \omega(k)\tilde{a}^\dagger_k \tilde{a}_k \notag \\  H_\text{int}&=\tilde{\gamma}\int_{0}^{}\mathrm{d}k \ \left(\sigma_+\tilde{a}_k+\tilde{a}^\dagger_k\sigma_-\right),
\end{align}
where $\omega(k)$ gives the waveguide dispersion relation and the operator $\tilde{a}_k$ destroys a photon of wavenumber $k$ while obeying $[\tilde{a}_k,\tilde{a}^\dagger_{k'}]=\delta\left(k-k'\right)$. We have assumed the fixed coupling rate $\tilde{\gamma}$ between optical modes of wavenumber $k$ and atomic transition and adopted the convention that unspecified lower and upper integration limits imply negative and positive infinity respectively. 

It is shown by e.g, Maier \cite{Maier2007} that the dispersion relation for waveguide confined optical modes is surface-plasmonic. We linearise this about some central wavenumber $k_0$ so that: $\tilde{\omega}(k)\approx\omega_0+v_g(k-k_0)$, where $v_g$ represents the photon group velocity. This means that
\begin{align}
{H}_0&=\frac{1}{2}\Omega\sigma_z+\int_{}^{}\mathrm{d}k \ \omega_0\tilde{a}^\dagger_k \tilde{a}_k+v_g(k-k_0) \tilde{a}^\dagger_k \tilde{a}_k \notag \\  H_\text{int}&=\tilde{\gamma}\int_{}^{}\mathrm{d}k \ \left(\sigma_+\tilde{a}_k+\tilde{a}^\dagger_k\sigma_-\right),
\end{align}
where we have also extended the limits of integration to cover the entirety of wavenumber space---an appropriate approximation when the band of populated modes is narrow. We next introduce the variable: $\epsilon\equiv v_g(k-k_0)$, which we use to re-write the Hamiltonian
\begin{align}
H_0&=\frac{1}{2}\Omega\sigma_z+\int\mathrm{d}\epsilon \ (\omega_0+\epsilon)a_\epsilon^\dagger a_\epsilon \notag \\  H_\text{int}&=\gamma\int\mathrm{d}\epsilon \ \left(\sigma_+a_\epsilon+\sigma_-a_\epsilon^\dagger\right) \label{eq:freqham}
\end{align}
where we have defined $\gamma\equiv v_g^{-\frac{1}{2}}\tilde{\gamma}$ and $a_\epsilon=v_g^{-\frac{1}{2}}\tilde{a}_{k_0+v_g^{-1}\epsilon}$. It can be easily shown that the commutation relation $[a_\epsilon,a_{\epsilon'}^\dagger]=\delta(\epsilon-\epsilon')$ is preserved. 

At this point we can simply use the definition of the interaction Hamiltonian \cite{schwartz2014quantum} and equation \eqref{eq:freqham} to deduce that
\begin{align}
H_\text{I}(t)=\gamma\int\mathrm{d}\epsilon \ (e^{-i\Delta_\epsilon t}\sigma_+a_\epsilon+e^{i\Delta_\epsilon t}\sigma_-a_\epsilon^\dagger), \label{eq:finhama}
\end{align}
which is the desired result, with the detuning defined by $\Delta_\epsilon\equiv\omega_0+\epsilon-\Omega$. Eq.~\eqref{eq:finhama} has the expected structure of an interaction Hamiltonian, with phases on the operators given by the energy mis-match between photons and emitter. The last point to note is the slight difference in structure between Hamiltonian \eqref{eq:finhama} and the version used by other authors (e.g. \cite{PhysRevA.82.063821}). The discrepancies can be ascribed simply to our not working in a frame rotating at the waveguide's central frequency and our inclusion of the free emitter Hamiltonian in $H_0$ as opposed to $H_\text{int}$. 

\section{Integration Technique}
\label{sec:int}
\noindent
For completeness we describe here the integration technique used to evaluate the explicit integral expressions for the single and two photon transition amplitudes. This is a relatively well-known result and can be found in e.g. the appendix of \cite{Branczykphdthesis}. We define the integral $\mathcal{I}$ and begin by changing variables so as to shift the limits of integration
\begin{align}
\mathcal{I}\equiv\int_{-\infty}^{t_1}\mathrm{d}t_2 \ e^{-i\Delta_it_2} \notag =\int_{0}^{\infty}\mathrm{d}t_2 \ e^{-i\Delta_i(t_1-t_2)}. \notag 
\end{align}
This can be decomposed and multiplied by unity to give
\begin{align}
\mathcal{I}=e^{-i\Delta_it_1}\lim\limits_{\alpha\rightarrow0}\int_{0}^{\infty}\mathrm{d}t_2 \ e^{-\alpha t_2}\left[ \cos\left(\Delta_it_2\right)+i \sin\left(\Delta_it_2\right)\right] 
\end{align}
and we then make use of standard results \cite{arfken2005mathematical}, for example noting
\begin{align}
\delta(x)=\frac{1}{\pi}\lim\limits_{a\rightarrow0}\frac{a}{a^2+x^2}
\end{align}
to find that
\begin{align}
\mathcal{I}&=e^{-i\Delta_it_1}\lim\limits_{\alpha\rightarrow0}\left(\frac{\alpha}{\alpha^2+\Delta_i^2}+i\frac{\Delta_{i}}{\alpha^2+\Delta_i^2}\right) \notag \\ &=e^{-i\Delta_it_1}\left(\pi\delta(\Delta_i)+\frac{i}{\Delta_i}\right), \label{eq:result}
\end{align}
which is the desired formula.

\section{Borel Summation}
\label{sec:Bor}
\noindent In order to find the single photon transition amplitude it is necessary to evaluate the sum
\begin{align}
\sigma=\sum_{n=1}^{\infty}\left(-\gamma^2\pi g(\Delta_i)\right)^n,
\end{align} 
which is rendered possible for the case of $\protect{\left|\pi\gamma^2g(\Delta_i)\right|<1}$ via the binomial theorem. Terms in the series are divergent when this condition is not satisfied and we therefore need to take a more nuanced approach to assign a value to the sum outside of this regime. In fact, such divergent series are a common occurrence in quantum electrodynamics \cite{PhysRev.85.631} and there are a range of methods used to extract meaning from them. The tool we utilise here is the \emph{Borel Summation}---a technique applied in a diverse range of fields \cite{kleinert2009path} to analyse series with the $n$\textsuperscript{th} term divergent up to a factor of $n!$. 

We first demand that $\Delta_i\neq0$ and find 
\begin{align}
\sigma=\sum_{n=0}^{\infty}\left(-\frac{i\pi\gamma^2}{\Delta_i}\right)^n.  \label{eq:sum}
\end{align}
The Borel transformation of Eq. \eqref{eq:sum} is defined by \cite{bender1978advanced}
\begin{align}
\phi(z)\equiv\sum_{n=0}^{\infty}\frac{1}{n!}(-i\pi z)^n=e^{-i\pi z}
\end{align}
and the Borel sum by 
\begin{align}
\mathcal{B}\left(\frac{\gamma^2}{\Delta_i}\right)\equiv\int_{0}^{\infty}\mathrm{d}t \ e^{-t}\phi\left(\frac{\gamma^2}{\Delta_i}t\right)=\frac{1}{1+i\frac{\pi\gamma^2}{\Delta_i}} \label{eq:sumres}
\end{align}
under the condition now that $\protect{\text{Im}\left[\frac{\pi\gamma^2}{\Delta_i}\right]<1}$. However, it is possible to derive the Heisenberg-Langevin equations associated with the Hamiltonian of Eq.~\eqref{eq:finham}, as in e.g. Ref.~\cite{PhysRevA.75.053823}, and in doing so we find that the emitter lifetime is directly proportional to $\gamma^{-2}$. This means that $\frac{\pi\gamma^2}{\Delta_i}$ is entirely real and thus the condition is always satisfied. The Borel-summed result is then
\begin{align}
\mathcal{A}=\delta(f-i)\frac{\Delta_i-i\pi \gamma^2}{\Delta_i+i\pi\gamma^2},
\end{align}
valid for all coupling strengths. 

\section{Equivalence to Fan Result}
\label{sec:Fan}
\noindent
In this section we demonstrate the equivalence between our results for the one and two-photon transition amplitudes for a TLS and those found by Fan \textit{et al}. \cite{PhysRevA.82.063821}. As our transition amplitudes are evaluated in the limit $t\rightarrow\infty$ and the single final atomic state $\ket{g}$ is assumed, then the scattering matrix is in fact the quantity given by these amplitudes. For the single photon case we find that
\begin{align}
\mathcal{A}=\frac{1-\pi\gamma^2g(\Delta_i)}{1+\pi\gamma^2g(\Delta_i)}\delta(f-i).
\label{eq:us}
\end{align}
We can substitute our definition of $g(\Delta)$ into Eq. \eqref{eq:us} to determine
\begin{align}
\mathcal{A}=\frac{\Delta_i-i\pi\gamma^2-\pi^2\gamma^2\Delta_i\delta(\Delta_i)}{\Delta_i+i\pi\gamma^2+\pi^2\gamma^2\Delta_i\delta(\Delta_i)}\delta(f-i)
\end{align}
which is naturally equal to that found by Fan \textit{et al}.
\begin{align}
\mathcal{A}=\frac{\Delta_i-i\pi\gamma^2}{\Delta_i+i\pi\gamma^2}\delta(f-i).
\end{align}

The two-photon result requires a little more effort, our result is that
\begin{widetext}
\begin{align}
\mathcal{A}=&[t(i_0)+t(i_1)-1][\delta(f_0-i_0)\delta(f_1-i_1)+\delta(f_0-i_1)\delta(f_1-i_0)] \notag\\ +&2\pi\gamma^4\delta(f_0+f_1-i_0-i_1)\sum_{s=0,1}^{}\sum_{s'=0,1} \frac{g(\Delta_{i_s})g(\Delta_{i_s}-\Delta_{f_{s'}})g(\Delta_{f_{s'\oplus1}})}{[1+\pi\gamma^2g(\Delta_{i_s})][1+\pi\gamma^2g(\Delta_{f_{s'\oplus1}})]}.
\end{align}
Now, if we expand out the factor $g(\Delta_{i_s}-\Delta_{f_{s'}})$ so that	
\begin{align}
\frac{g(\Delta_{i_s})g(\Delta_{i_s}-\Delta_{f_{s'}})g(\Delta_{f_{s'\oplus1}})}{[1+\pi\gamma^2g(\Delta_{i_s})][1+\pi\gamma^2g(\Delta_{f_{s'\oplus1}})]}=&  \frac{\pi g(\Delta_{i_s})\delta(\Delta_{i_s}-\Delta_{f_{s'}})g(\Delta_{f_{s'\oplus1}})}{[1+\pi\gamma^2g(\Delta_{i_s})][1+\pi\gamma^2g(\Delta_{f_{s'\oplus1}})]} \notag \\&+\frac{ig(\Delta_{i_s})g(\Delta_{f_{s'\oplus1}})}{(\Delta_{i_s}-\Delta_{f_{s'}})[1+\pi\gamma^2g(\Delta_{i_s})][1+\pi\gamma^2g(\Delta_{f_{s'\oplus1}})]}.
\end{align}
It is then true that
	\begin{align}
	\mathcal{A}=&[t(i_0)+t(i_1)-1][\delta(f_0-i_0)\delta(f_1-i_1)+\delta(f_0-i_1)\delta(f_1-i_0)] \notag\\ &+4\pi^2\gamma^4\left[\delta(f_0-i_0)\delta(f_1-i_1)+\delta(f_0-i_1)\delta(f_1-i_0)\right]\frac{g(\Delta_{i_0})g(\Delta_{i_1})}{\left[1+\pi\gamma^2g(\Delta_{i_0})\right]\left[1+\pi\gamma^2g(\Delta_{i_1})\right]} \notag \\ &+2\pi i\gamma^4\delta(f_0+f_1-i_0-i_1)\left[\frac{g(\Delta_{i_0})g(\Delta_{f_0})}{(\Delta_{i_0}-\Delta_{f_1})\left[1+\pi\gamma^2g(\Delta_{i_0})\right]\left[1+\pi\gamma^2g(\Delta_{f_0})\right]}\right. \notag \\ &+\left. \frac{g(\Delta_{i_0})g(\Delta_{f_1})}{(\Delta_{i_0}-\Delta_{f_0})\left[1+\pi\gamma^2g(\Delta_{i_0})\right]\left[1+\pi\gamma^2g(\Delta_{f_1})\right]} +\frac{g(\Delta_{i_0})g(\Delta_{f_1})}{(\Delta_{i_1}-\Delta_{f_0})\left[1+\pi\gamma^2g(\Delta_{i_1})\right]\left[1+\pi\gamma^2g(\Delta_{f_1})\right]}\right. \notag \\ &+\left.\frac{g(\Delta_{i_1})g(\Delta_{f_0})}{(\Delta_{i_1}-\Delta_{f_1})\left[1+\pi\gamma^2g(\Delta_{i_1})\right]\left[1+\pi\gamma^2g(\Delta_{f_0})\right]}  \right].
	\end{align}
	We can rearrange the frequency-conserving terms and again substitute the definition of $g(\Delta)$ into the frequency-mixing term to determine
	\begin{align}
	\mathcal{A}=t(i_0)t(i_1)[\delta(f_0-i_0)\delta(f_1-i_1)&+\delta(f_0-i_1)\delta(f_1-i_0)] \notag\\  +\frac{2\pi i\gamma^4\delta(f_0+f_1-i_0-i_1)}{\left[\Delta_{f_0}+i\pi\gamma^2\right]\left[\Delta_{f_1}+i\pi\gamma^2\right]} & \left[\frac{\Delta_{f_1}+i\pi\gamma^2}{(\Delta_{f_1}-\Delta_{i_0})\left[\Delta_{i_0}+i\pi\gamma^2\right]}+ \frac{\Delta_{f_0}+i\pi\gamma^2}{(\Delta_{f_0}-\Delta_{i_0})\left[\Delta_{i_0}+i\pi\gamma^2\right]} \right. \notag \\ &+\left.\frac{\Delta_{f_0}+i\pi\gamma^2}{(\Delta_{f_0}-\Delta_{i_1})\left[\Delta_{i_1}+i\pi\gamma^2\right]}+\frac{\Delta_{f_1}+i\pi\gamma^2}{(\Delta_{f_1}-\Delta_{i_1})\left[\Delta_{i_1}+i\pi\gamma^2\right]}  \right]. \label{eq:near}
	\end{align}
Straight-forward algebraic manipulation of Eq. \eqref{eq:near} then leads us to
\begin{align}
\mathcal{A}&=t(i_0)t(i_1)[\delta(f_0-i_0)\delta(f_1-i_1)+\delta(f_0-i_1)\delta(f_1-i_0)] +\frac{4\pi i\gamma^4\delta(f_0+f_1-i_0-i_1)}{\left[\Delta_{f_0}+i\pi\gamma^2\right]\left[\Delta_{f_1}+i\pi\gamma^2\right]} \left(\frac{1}{\Delta_{i_0}+i\pi\gamma^2}+\frac{1}{\Delta_{i_1}+i\pi\gamma^2}\right)
\end{align}
which is the result by Fan \textit{et al}. 
\end{widetext}

\section{Vanishing Terms in Higher Order Transition Amplitudes}

\label{sec:zer}
\begin{figure}[t!]
	\begin{center}
		\includegraphics[scale=1]{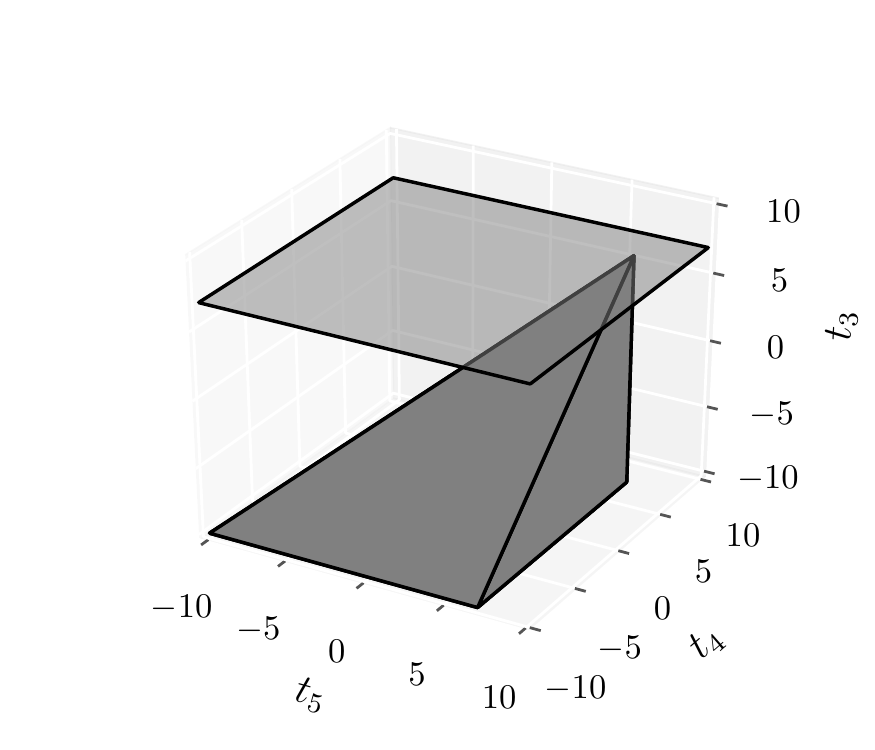}
		\caption{Integration region defined by the enclosed volume. We see that it intersects the surface defined by $t_5=t_2$ at only a single point.}
		\label{fig:intreg}
	\end{center}
\end{figure}
\noindent
In this appendix we show mathematically why diagrams with internal photon loops spanning multiple time integrals, as described in Sec. \ref{sec:tp} and shown in Fig. \ref{fig:forbid}, should be discarded. If we methodically calculate $\mathcal{A}^{(6)}$ we arrive upon many terms, for example
\begin{align}
-\gamma^6\int\mathrm{d}\tilde{t}^{(6)} \int\mathrm{d}\omega \ e^{i(\Delta_{f_1}t_1-\Delta_\omega (t_2-t_5)+\Delta_{f_0}t_3-\Delta_{i_0}t_4-\Delta_{i_1}t_6)}.
\end{align}
Evaluation of the frequency integral in this expression yields the Dirac delta function $\delta(t_5-t_2)$ and so we are evaluating an integral of the form
\begin{align}
\int_{}^{t_2}\mathrm{d}t_3\int_{}^{t_3}\mathrm{d}t_4\int_{}^{t_4}\mathrm{d}t_5 \ h(t_5,t_4,t_3) \delta(t_5-t_2),
\end{align}
where $h(t_5,t_4,t_3)$ is some exponential function. The integral here is over a volume in time-space, bounded by the surfaces $t_5=t_4$ and $t_4=t_3$. The delta function has the effect of converting this volume integral into one over a surface---where the surface is defined by projection of the original volume onto $t_5=t_2$. A representation of this is depicted in Figure \ref{fig:intreg} and we see that the resulting surface is given by a point. This term therefore does not contribute to the transition amplitude.

\section{Integral and Diagrammatic Evaluation of $\mathcal{A}^{(8)}$ for the Two Photon Case}
\label{sec:exp}
\subsection{Direct Integration Approach}
\noindent In this appendix we demonstrate that for $n=8$ the diagrammatic and integral approaches to evaluation of the $n$\textsuperscript{th} order transition amplitude agree. By definition we have that
\begin{widetext}
	\begin{align}
	\mathcal{A}^{(8)}=\gamma^8\int\mathrm{d}\tilde{t}^{(8)}\int\mathrm{d}\bar{\epsilon}^{(8)} \ e^{i(\Delta_{\epsilon_1}t_1-\Delta_{\epsilon_2}t_2+\Delta_{\epsilon_3}t_3-\Delta_{\epsilon_4}t_4+\Delta_{\epsilon_5}t_5-\Delta_{\epsilon_6}t_6+\Delta_{\epsilon_7}t_7-\Delta_{\epsilon_8}t_8)}& \notag \\ \times \bra{0}a_{f_0}a_{f_1}a_{\epsilon_1}^\dagger a_{\epsilon_2}a_{\epsilon_3}^\dagger a_{\epsilon_4}a_{\epsilon_5}^\dagger a_{\epsilon_6}a_{\epsilon_7}^\dagger a_{\epsilon_8}a_{i_{1}}^\dagger a_{i_0}^\dagger\ket{0}&.
	\end{align}
	The vacuum-expectation-value in this expression can be directly evaluated and we find expressions for a total of thirty-two terms
	\begin{align}
	\mathcal{A}^{(8)}=\gamma^8\int\mathrm{d}\tilde{t}^{(8)}\int\mathrm{d}\epsilon_1\int\mathrm{d}\epsilon_2 \  \Big[&e^{i(\Delta_{f_0}t_1-\Delta_{\epsilon_1}t_2+\Delta_{f_1}t_3-\Delta_{\epsilon_2}t_4+\Delta_{\epsilon_1}t_5-\Delta_{i_1}t_6+\Delta_{\epsilon_2}t_7-\Delta_{i_0}t_8)} \notag \\   +&e^{i(\Delta_{f_1}t_1-\Delta_{\epsilon_1}t_2+\Delta_{f_0}t_3-\Delta_{\epsilon_2}t_4+\Delta_{\epsilon_1}t_5-\Delta_{i_1}t_6+\Delta_{\epsilon_2}t_7-\Delta_{i_0}t_8)}  \notag \\ +&\ldots  \notag \\  +&e^{i(\Delta_{f_1}t_1-\Delta_{i_0}t_2+\Delta_{f_0}t_3-\Delta_{\epsilon_2}t_4+\Delta_{\epsilon_2}t_5-\Delta_{\epsilon_1}t_6+\Delta_{\epsilon_1}t_7-\Delta_{i_1}t_8)} \Big]
	\end{align}
\end{widetext}
where we have used the delta functions from the decomposed vacuum-expectation-value to eliminate six of the eight frequency integrals. We can then use the definition of the Dirac delta function to transform the remaining frequency integrals and integrands into delta functions in time. Using the method outlined in App. \ref{sec:zer}, we can then eliminate any term with a delta function connecting non-adjacent times (e.g. $\delta(t_7-t_4)$, $\delta(t_4-t_1)$ etc.) and sixteen terms remain. There are however only four `categories' of term---with each category containing four terms that are permutations over initial and final photon energies. We find that
\begin{widetext}
\begin{align}
\mathcal{A}^{(8)}&=(2\pi)^2\gamma^8\sum_{s=0,1}^{}\sum_{s'=0,1}^{}\int\mathrm{d}\tilde{t}^{(8)} \notag \Big[2\pi\delta(f_{s'}-i_s)\delta(t_7-t_6)\delta(t_5-t_4)\delta(t_3-t_2) \ e^{i(\Delta_{f_{s'\oplus1}}t_1-\Delta_{i_{s\oplus1}}t_8)}  \notag \\ + &\delta(t_7-t_6)\delta(t_5-t_4)e^{i(\Delta_{f_{s'}}t_1-\Delta_{i_{s\oplus1}}t_2+\Delta_{f_{s'\oplus1}}t_3-\Delta_{i_{s}}t_8)} + \delta(t_3-t_2)\delta(t_7-t_6)e^{i(\Delta_{f_{s'}}t_1-\Delta_{i_{s\oplus1}}t_4+\Delta_{f_{s'\oplus1}}t_5-\Delta_{i_{s}}t_8)}  \notag \\ +&\delta(t_5-t_4)\delta(t_3-t_2)e^{i(\Delta_{f_{s'}}t_1-\Delta_{i_{s\oplus1}}t_6+\Delta_{f_{s'\oplus1}}t_7-\Delta_{i_{s}t_8})}\Big] \label{eq:a8} .
\end{align}
The integrals in Eq. \eqref{eq:a8} can be evaluated directly, as in the main text for $n=6$ and we find
\begin{align}
\mathcal{A}^{(8)}&=2\pi^3\gamma^8\sum_{s=0,1}^{}\sum_{s'=0,1}^{}  \Big[\pi g^4(\Delta_{i_s})\delta(f_{s'}-i_s)\delta(f_{s'\oplus1}-i_{s\oplus1})   + g^3(\Delta_{i_s})g(\Delta_{i_s}-\Delta_{f_{s'}})g(\Delta_{f_{s'\oplus1}})\delta(f_0+f_1-i_0-i_1) \notag \\ + &g^2(\Delta_{i_s})g(\Delta_{i_s}-\Delta_{f_{s'}})g^2(\Delta_{f_{s'\oplus1}})\delta(f_0+f_1-i_0-i_1)    +g(\Delta_{i_s})g(\Delta_{i_s}-\Delta_{f_{s'}})g^3(\Delta_{f_{s'\oplus1}})\delta(f_0+f_1-i_0-i_1) \Big] \label{eq:a8n} ,
\end{align}
\end{widetext}
which is the final result for the $n=8$ term in the two-photon transition amplitude. 
\subsection{Diagrammatic Method}
\noindent	 
The diagrams corresponding to the four species of term in Eqn. \eqref{eq:a8} are shown in Fig. \ref{fig:diagrams}. In Fig. \ref{fig:nomix} a single photon is absorbed and emitted by the atom four times, in Fig. \ref{fig:b} a photon is absorbed and emitted three times, before a second photon is absorbed and emitted once. Fig. \ref{fig:c} shows both photons being absorbed and emitted twice and the Fig. \ref{fig:d} shows a single absorption/emission for the first photon, followed by three for the second. Diagram \ref{fig:nomix} represents the non-frequency mixing component of the $n=8$ term and therefore contributes a factor of $2\pi^4\gamma^8g^4(\Delta_{i_s})\delta(f_{s'}-i_{s})\delta(f_{s'\oplus1}-i_{s\oplus1})$ to the amplitude---this being the single photon result multiplied by an additional delta function to impose conservation of energy for the second photon.

The three frequency mixing diagrams require application of the rules supplied in the main text. For example, consider the diagram shown in Fig. \ref{fig:diagrams}\textcolor{red}{c}. We first associate the pre-factor $2\pi^3\gamma^8$ to this diagram's term, substituting $n=8$ into the expression $\frac{2}{\pi}(\sqrt{\pi}\gamma)^n$ for the $n$\textsuperscript{th} order case. The first absorption event then yields a factor of $g(\Delta_{i_s})$ as per the rules and we gain an additional factor of this term from the internal emission and absorption of the $\epsilon_1$ photon. Emission of the photon with frequency $f_{s'}$ then yields the factor $g(\Delta_{i_s}-\Delta_{f_{s'}})$ before the next incident photon is absorbed, producing $g(\Delta_{i_{s\oplus1}}+\Delta_{i_s}-\Delta_{f_{s'}})$. One additional copy of this factor is required, because of the second internal photon emission/absorption process but its argument can be simplified, as the final emission event yields the factor $\delta(f_0+f_1-i_0-i_1)$, meaning that $\Delta_{i_{s\oplus1}}+\Delta_{i_s}-\Delta_{f_{s'}}=\Delta_{f_{s'\oplus1}}$. Multiplying individual factors together yields the expression $2\pi^3\gamma^8g^2(\Delta_{i_s})g(\Delta_{i_s}-\Delta_{f_{s'}})g^2(\Delta_{f_{s'\oplus1}})\delta(f_0+f_1-i_0-i_1)$, exactly as found in Eq. \eqref{eq:a8n}. 

\section{Equivalence of $\mathcal{A}_{\mu\nu}$ and $T^{(2)}(\omega)$}
\label{app:njp}
\noindent
In Ref.~\cite{1367-2630-14-9-095028}, Pletyukhov and Gritsev derive the following expression (their Eq.~(46)) for the $T$-matrix when two photons scatter from a $\Lambda$-system
\begin{align}
T^{(2)}(\omega)=&\left[g_{31}^2P_1+g_{32}^2P_2+g_{31}g_{32}\left(\ket{2}\!\bra{1}+\ket{1}\!\bra{2}\right)\right]\notag \\ &\times a_{\nu_1}^\dagger G_3a_{\nu_2}\left(g_{31}^2G_1+g_{32}^2G_2\right)a_{\nu_3}^\dagger G_3a_{\nu_4}. \label{eq:njpres}
\end{align}
Here $g_{31}$ and $g_{32}$ represent the ground-excited state couplings, equivalent to our $\gamma_1$ and $\gamma_2$ respectively. The states $\ket{1}$ and $\ket{2}$ are the atomic ground levels, which we labelled $\ket{g_1}$ and $\ket{g_2}$ and the operators $P_1$ and $P_2$ project onto these states. Bosonic operators are given by $a_\nu$ and propagators by $G_{1/2/3}$. 

The first step in demonstrating the equivalence between Eq.~\eqref{eq:njpres} and our Eq.~\eqref{eq:finaltpl} is to apply the `intertwining property' (Eq.~(15) in Ref.~\cite{1367-2630-14-9-095028}) and rearrange bosonic operators and propagators. This has the effect of adding terms to the propagator's denominator and we note this in the propagator argument. Note that integration over the internal photon frequencies $\nu$ is implied and we arrive upon
\begin{widetext}
\begin{align}
T^{(2)}(\omega)=\left[g_{31}^2P_1+g_{32}^2P_2+g_{31}g_{32}\left(\ket{2}\!\bra{1}+\ket{1}\!\bra{2}\right)\right]G_3(\nu_1)&\left[g_{31}^2G_1(\nu_1-\nu_2)  +g_{32}^2G_2(\nu_1-\nu_2)\right] \notag \\ &G_3(\nu_1+\nu_3-\nu_2)a_{\nu_1}^\dagger a_{\nu_2}a_{\nu_3}^\dagger a_{\nu_4}. \label{eq:sandwiched}
\end{align}
\end{widetext}
Applying Wick's Theorem to the string of bosonic operators on the right hand side of Eq.~\eqref{eq:sandwiched}, we see that there are two distinct species of term in the resulting $T$-matrix elements. The first is where the `internal' photon annihilation operators are contracted only with `external' photon creation operators, which we label with frequencies $k_0$ and $k_1$. In this case energies of incoming and outgoing particles are not individually preserved. In the second species of term, the operator $a_{\nu_3}^\dagger$ is contracted with $a_{\nu_2}$. This means that an additional delta function arises from the overlap between incoming and outgoing states.

Using the definition of the $T$-matrix in the two-photon sector, we determine that the requirement for equivalence between our result and that of Pletyukhov and Gritsev is
\begin{align}
-2\pi i T_{p_0,p_1,k_0,k_1}^{(2)\mu\nu}\delta(E^\text{out}-E^\text{in})\stackrel{?}{=}\mathcal{A}_{\mu\nu}-\mathcal{N}_{\mu\nu} \label{eq:cond},
\end{align}
where $T^{(2)\mu\nu}_{p_0,p_1,k_0,k_1}\equiv \bra{p_0,p_1;\mu}T^{(2)}\ket{k_0,k_1;\nu}$ and $T^{(2)}$ is defined as the on-shell evaluation of $T^{(2)}(\omega)$.  We see immediately that the bound-state contributions to the left (LHS) and right hand side (RHS) of Eq.~\eqref{eq:cond} are indeed equivalent. Consider
\begin{widetext}
\begin{align}
-2\pi i g_{3\mu}g_{3\nu}g_{31}^2G_3^\text{os}G_1^\text{os}G_3^\text{os}\delta_{\nu_2k_1}\delta_{\nu_4k_0}\delta_{\nu_1p_1}\delta_{\nu_3p_0}&= \frac{1}{g_{3\mu}g_{3\nu}}\frac{-2\pi ig_{31}^2}{k_0+\varepsilon_\nu-p_0-\varepsilon_1}\frac{g_{3\mu}g_{3\nu}}{p_1+\varepsilon_\mu-\varepsilon_3+i\pi g^2}\frac{g_{3\mu}g_{3\nu}}{k_0+\varepsilon_\nu-\varepsilon_3+i\pi g^2} \notag \\ &=\frac{i}{2\pi \gamma_{\mu}\gamma_{\nu}}\frac{1}{k_0+\varepsilon_\nu-p_0-\varepsilon_1}\frac{2\pi i \gamma_\mu\gamma_\nu}{\Delta_{p_1\mu}+i\pi\gamma^2}\frac{2\pi i\gamma_\mu\gamma_\nu}{\Delta_{k_0\nu}+i\pi \gamma^2}, \label{eq:njpmixed}
\end{align}
\end{widetext}
where $G^\text{os}$ indicates that the propagator is to be evaluated on-shell, we made the transformations $g_{3i}\rightarrow\gamma_i$, $\varepsilon_3\rightarrow\Omega$ and defined $\gamma^2\equiv\gamma_1^2+\gamma_2^2$. We see that Eq.~\eqref{eq:njpmixed} is nothing more than one of the eight components of the bound state amplitude in Eq.~\eqref{eq:finaltpl}. As we must sum over all possible permutations of initial and final photon states (and both of the propagators $G_1$ and $G_2$), the proof is complete.

It then only remains to treat the component of Eq.~\eqref{eq:cond} where individual photon energies are preserved separately. Consider
\begin{widetext}
\begin{align}
-2\pi^2 g_{3\mu}g_{3\nu}g_{31}^2G^\text{os}_3G^\text{os}_3\delta_{\nu_2\nu_3}\delta_{\nu_4k_0}\delta_{\nu_1p_0}\delta_{p_1+\mu,k_1+\varepsilon_1}=\frac{\gamma_1^2}{2\gamma_\mu\gamma_\nu}\frac{2\pi i\gamma_\mu\gamma_\nu}{\Delta_{k_0\nu}+i\pi g^2}\frac{2\pi i\gamma_\mu\gamma_\nu}{\Delta_{p_0\mu}+i\pi g^2}\delta_{p_1+\mu,k_1+\varepsilon_1}, \label{eq:njppres}
\end{align}
\end{widetext}
where we have simply made the same transformations as for the bound-state case and also applied Eq.~(16) of Ref.~\cite{1367-2630-14-9-095028}. Again, Eq.~\eqref{eq:njppres} is one of the eight components of the non-bound state portion of Eq.~\eqref{eq:finaltpl} and, owing to the sum over internal propagators, initial and final photon configurations, our main result is recovered.

\end{document}